
\input phyzzx
\catcode`\@=11
\paperfootline={\hss\iffrontpage\else\ifp@genum\tenrm
 -- \folio\ --\hss\fi\fi}
\def\titlestyle#1{\par\begingroup \titleparagraphs
 \iftwelv@\fourteenpoint\fourteenbf\else\twelvepoint\twelvebf\fi
 \noindent #1\par\endgroup }
\def\GENITEM#1;#2{\par \hangafter=0 \hangindent=#1
 \Textindent{#2}\ignorespaces}
\def\address#1{\par\kern 5pt\titlestyle{\twelvepoint\sl #1}}
\def\abstract{\par\dimen@=\prevdepth \hrule height\z@
\prevdepth=\dimen@
 \vskip\frontpageskip\centerline{\fourteencp Abstract}\vskip\headskip }
\newif\ifYUKAWA  \YUKAWAtrue
\font\elevenmib   =cmmib10 scaled\magstephalf
\skewchar\elevenmib='177
\def\YUKAWAmark{\hbox{\elevenmib
 Yukawa\hskip0.05cm Institute\hskip0.05cm Kyoto \hfill}}
\def\titlepage{\FRONTPAGE\papers\ifPhysRev\PH@SR@V\fi
 \ifYUKAWA\null\vskip-1.70cm\YUKAWAmark\vskip0.6cm\fi
 \ifp@bblock\p@bblock \else\hrule height\z@ \rel@x \fi }

\def\schapter#1{\par \penalty-300 \vskip\chapterskip
 \spacecheck\chapterminspace
 \chapterreset \titlestyle{\ifcn@@\S\ \chapterlabel.~\fi #1}
 \nobreak\vskip\headskip \penalty 30000
 {\pr@tect\wlog{\string\chapter\space \chapterlabel}} }

\def\ssection#1{\par \ifnum\lastpenalty=30000\else
 \penalty-200\vskip\sectionskip \spacecheck\sectionminspace\fi
 \gl@bal\advance\sectionnumber by 1
 {\pr@tect
 \xdef\sectionlabel{\ifcn@@ \chapterlabel.\fi
 \the\sectionstyle{\the\sectionnumber}}%
 \wlog{\string\section\space \sectionlabel}}%
 \noindent {\S \caps\thinspace\sectionlabel.~~#1}\par
 \nobreak\vskip\headskip \penalty 30000 }


\papers

\def\lkakko{\vbox{\vskip0.065cm\hbox{(}\vskip-0.065cm}}
\def\rkakko{\vbox{\vskip0.065cm\hbox{)}\vskip-0.065cm}}
\def\YUKAWAHALL{\hbox to \hsize
 {\hfil \lkakko\twelvebf YUKAWA HALL\rkakko\hfil}}


\def\Endline{\hfil\break}


\def\addeqno{\ifnum\equanumber<0 \global\advance\equanumber by -1
 \else \global\advance\equanumber by 1\fi}


\mathchardef\Lag="724C
\def\sqr#1#2{{\vcenter{\hrule height.#2pt
 \hbox{\vrule width.#2pt height#1pt \kern#1pt\vrule width.#2pt}
 \hrule height.#2pt}}}


\def\cref#1{\rlap,\attach{#1)}}
\def\ref#1{\attach{#1)}}



\newdimen\ex@
\ex@.2326ex
\def\boxed#1{\setbox\z@\hbox{$\displaystyle{#1}$}\hbox{\lower.4\ex@
 \hbox{\lower3\ex@\hbox{\lower\dp\z@\hbox{\vbox{\hrule height.4\ex@
 \hbox{\vrule
width.4\ex@\hskip3\ex@\vbox{\vskip3\ex@\box\z@\vskip3\ex@}%
 \hskip3\ex@\vrule width.4\ex@}\hrule height.4\ex@}}}}}}
\def\txtboxed#1{\setbox\z@\hbox{{#1}}\hbox{\lower.4\ex@
 \hbox{\lower3\ex@\hbox{\lower\dp\z@\hbox{\vbox{\hrule height.4\ex@
 \hbox{\vrule
width.4\ex@\hskip3\ex@\vbox{\vskip3\ex@\box\z@\vskip3\ex@}%
 \hskip3\ex@\vrule width.4\ex@}\hrule height.4\ex@}}}}}}
\newdimen\exx@
\exx@.1ex
\def\thinboxed#1{\setbox\z@\hbox{$\displaystyle{#1}$}\hbox{\lower.4\exx@
 \hbox{\lower3\exx@\hbox{\lower\dp\z@\hbox{\vbox{\hrule height.4\exx@
 \hbox{\vrule width.4\exx@\hskip3\exx@%
 \vbox{\vskip3\ex@\box\z@\vskip3\exx@}%
 \hskip3\exx@\vrule width.4\exx@}\hrule height.4\exx@}}}}}}

\chardef\fontD="1A

\catcode`@=12

%
\pubnum={YITP/K-1079}
\date={July 1994}

\titlepage
\title{CANONICAL 3+1 DESCRIPTION OF RELATIVISTIC MEMBRANES}

\author{Jens Hoppe\footnote*{
Heisenberg Fellow\Endline
On leave of absence from the Institute for Theoretical Physics,
Karlsruhe University.}
}

\address{
Yukawa Institute for Theoretical Physics\break
Kyoto University,~Kyoto 606,~Japan}

\abstract{
$M$-dimensional extended objects $\Sigma$ can be described
by projecting a Diff $\Sigma$ invariant Hamiltonian of
time-independent Hamiltonian density {\cal H} onto the
Diff $\Sigma$- singlet sector, which after Hamiltonian reduction,
using {\cal H} itself for one of the gauge-fixing conditions,
results in a non-local description that may enable one to
extend the non-local symmetries for strings to higher dimensions and
make contact with gravity at an early stage.
}

\endpage

Whereas the Hamiltonian light-cone description of relativistic
membranes
(cf.[1]) has widely been used [2], a corresponding $3+1$ formulation,
that
would e.g. provide a Hamiltonian structure for the 'generalized
$su(\infty )-$ Nahm-equations' (derived in [3]) as well as the steady
state
irrotational isentropic inviscid K\'arm\'an-Tsien gas (see [4]),
appears to be missing.
Filling this gap turns out to reveal a number of rather interesting
features of membrane theory ( more generally, the theory of massless
extended objects of arbitrary dimension ). The disappearance of one of
the light-cone coordinates from the light-cone Hamiltonian, e.g., finds
its correspondence in the time-independence of the Hamiltonian
density(!)
in the $`n+1'$-formulation ( which can then be used to partly fix the
invariance under time-independent reparametrisations of the extended
object ). The Hamiltonian equations of motion can be shown to be
implied
by n infinite sets of conservation laws. Though the complete
Hamiltonian
reduction is difficult to perform explicitly, it is likely that
a non-local description similar to the loop-representation
in  general relativity ( see e.g. [5] ) will result.
As most of the considerations apply to general $M$-dimensional extended
objects    moving in $M+1$ dimensional euclidean space ($M=2$ for
membranes),
I will start by considering the relativistic minimal hyper-surface-
problem in arbitrary space-time dimensions $D$, i.e. embeddings of
$n$-dimensional manifolds ${\cal M}$ ( of signature ($+,~-,~\cdots,~-$)
)
into $n+1$-dimensional Minkowski-space, for which the first variation
of their volume $S$,
$$
\eqalign{
& S = \int_{{\cal M}} d^n \varphi  \sqrt{G} \cr
& G = (-)^M det \big( {\partial x^\mu  \over \partial \varphi^\alpha }
{}~
{\partial x^\nu  \over \partial \varphi^\beta } ~
\eta _{\mu \nu } \big)_{\alpha , \beta  = 0,\hbox to 3mm{} \cdots
M=n-1} \cr
& \eta _{\mu \nu } = diag ( 1, ~-1,~\cdots,~-1)~~, \cr
}
\eqno{(1)}
$$
vanishes.
In order to simplify the equations of motion,
$$
\eqalign{
{1 \over \sqrt{G}}~\partial_\alpha  \sqrt{G}~G^{\alpha \beta }
\partial_\beta ~x^\mu  = 0
 &\hbox to 10mm{} \mu  =0~\cdots~n~~, \cr
}
\eqno{(2)}
$$

one may choose $\varphi^0$ to be the time $t = x^0$
( leaving the timedependent shape of $\Sigma$,
$\displaystyle\mathop{x}^\rightharpoonup = (x^1,~\cdots, ~x^n) =
\mathop{x}^\rightharpoonup (t,~\varphi^1,~\cdots, \varphi^M)$
to be determined ), and use the remaining invariance under
timedependent
spatial reparametrisations,
$\varphi^r \rightarrow \tilde{\varphi}^r (t,~\varphi^1 \cdots
\varphi^M)$
to demand
$$
\dot{\mathop{x}^\rightharpoonup} \partial_r \mathop{x}^\rightharpoonup
= 0~~, \hbox to 10mm{} r=1, \cdots, M~~.
\eqno{(3)}
$$
The special feature of a hypersurface is that the $\mu  = i = 1$,
$\cdots,~n = M+1$ part of (2) is automatically satisfied,
provided (3), and the $\mu  = 0$ part of (2),
$$
\partial_t
\left( {g \over\displaystyle{1- \dot{\mathop{x}^\rightharpoonup}^2}}
\right)^{1/2}
= 0
\eqno{(4)}
$$
holds; $g$ denotes the determinant of the MxM matrix formed by
$g_{rs}:$ $= \partial_r \displaystyle\mathop{x}^\rightharpoonup
\partial_s \displaystyle\mathop{x}^\rightharpoonup$.\break
Minimal $n$ dimensional hypersurfaces can therefore be described by
the $n$ first-order equations(4), once integrated, and (3).
These may be put into Hamiltonian form by restricting the Diff
$\Sigma $ invariant Hamiltonian
$$
H = \int _\Sigma d^M \varphi \sqrt{{\mathop{p}^\rightharpoonup}^2 + g}
\eqno{(5)}
$$
to the 'singlet-sector',
$$
C_r: =
{\mathop{p}^\rightharpoonup}~{\displaystyle{\partial_r
{\mathop{x}^\rightharpoonup}}%
\over{\sqrt{{\displaystyle{\mathop{p}^\rightharpoonup}}^2 + g }}} = 0
{}~~.
\eqno{(6)}
$$
To check this, one first makes sure that, due to the canonical
equations
of motion,
$$
\dot{\mathop{x}^\rightharpoonup} =
{{\displaystyle{\mathop{p}^\rightharpoonup}} \over%
{\sqrt{ \displaystyle{\mathop{p}^\rightharpoonup}^2 + g }}}~~, \hbox to
5mm{}
\dot{\mathop{p}^\rightharpoonup} =
\partial_r \left( {{gg^{rs}\partial_s
\displaystyle{\mathop{x}^\rightharpoonup}}\over%
{\sqrt{ {\displaystyle{\mathop{p}^\rightharpoonup}}^2 + g } }}
\right)~~,
\eqno{(7)}
$$
$C_r$ is time-independent. Using (6), one then finds that
the Hamiltonian density ${\cal H}$ is also(!) conserved:
$$
\partial_t \left( \sqrt{ \displaystyle{\mathop{p}^\rightharpoonup}^2 +
g} \right) = 0~~.
\eqno{(8)}
$$
As before, the second order equations for
$\displaystyle\mathop{x}^\rightharpoonup$
are then automatically satisfied:
$$
\ddot{\mathop{x}^\rightharpoonup} =
{1 \over {\cal H}}~\partial_r
\left( {{gg^{rs}\partial_s
\displaystyle\mathop{x}^\rightharpoonup}\over%
{\cal H}} \right)~~.
\eqno{(9)}
$$
Note that (6) and (8) ( hence (9) ) is also consistent with the
equations
of motion derived by choosing as Hamiltonian density any non-linear
function of ${\cal H}$ ( in these cases, however, the equations of
motion alone
will not be sufficient to make
$\displaystyle\mathop{p}^\rightharpoonup \cdot \partial_r
\displaystyle\mathop{x}^\rightharpoonup$
proportional to some conserved quantity ).

The Hamiltonian form (5)/(6) may be derived in a less ad hoc way,
by the standard canonical procedure, just choosing $\varphi^0 = x^0$
( and $-S$ instead of $S$ ), leaving
$$
G_{\alpha \beta }~=~\pmatrix{
1- \displaystyle\dot{\mathop{x}^\rightharpoonup}^2  &
- \displaystyle\dot{\mathop{x}^\rightharpoonup}
\partial_r~{\mathop{x}^\rightharpoonup} \cr
- \displaystyle\dot{\mathop{x}^\rightharpoonup}
\partial_r~{\mathop{x}^\rightharpoonup} &
-g_{rs} \cr}~~,
\eqno{(10)}
$$
$$
{\cal L} = -\sqrt{G} = -\sqrt{g}~
\sqrt{(1-\dot{\mathop{x}^\rightharpoonup}^2) +
(\dot{\mathop{x}^\rightharpoonup} \partial_r
\mathop{x}^\rightharpoonup)%
{}~g^{rs}~(\dot{\mathop{x}^\rightharpoonup} \partial_s
\mathop{x}^\rightharpoonup)
}~~.
\eqno{(11)}
$$
Defining canonical momenta,
$$
p_i = {{\delta {\cal L}}\over{\delta \dot{x}_i}} =
\sqrt{
{g\over
{(1-\displaystyle\dot{\mathop{x}^\rightharpoonup}^2) +
(\displaystyle\dot{\mathop{x}^\rightharpoonup} \partial_r
\displaystyle\mathop{x}^\rightharpoonup)%
{}~g^{rs}~(\displaystyle\dot{\mathop{x}^\rightharpoonup}
\partial_s \displaystyle\mathop{x}^\rightharpoonup})}}
(\dot{x}_i - \partial_r x_i g^{rs} (\dot{\mathop{x}^\rightharpoonup}
\partial_s \mathop{x}^\rightharpoonup))~~,
\eqno{(12)}
$$
it is easy to see that
$$
{\cal H}: = \dot{\mathop{x}^\rightharpoonup}~\cdot~
\mathop{p}^\rightharpoonup - {\cal L} =
\sqrt{{\mathop{p}^\rightharpoonup}^2 + g}
\eqno{(13)}
$$
and
$$
\phi _r: = \mathop{p}^\rightharpoonup \partial_r
\mathop{x}^\rightharpoonup
\equiv 0~~, ~~~~r=1, \cdots, M
\eqno{(14)}
$$
( as a consequence of (12), i.e. without assuming (3) ).
The $\phi _r$
are primary first class constraints ( their Poissonbrackets among
themselves, and with ${\cal H}$ vanish on the constraint surface ).
According to Dirac [6], one should use
$$
H_T : = \int_\Sigma \sqrt{ {\mathop{p}^\rightharpoonup}^2 + g }
+ \int_\Sigma u^r \phi _r~~,
\eqno{(15)}
$$
leading to the equations of motion
$$
\eqalign{
\dot{\mathop{x}^\rightharpoonup} &
= {{\displaystyle\mathop{p}^\rightharpoonup}\over%
{\sqrt{ \displaystyle{\mathop{p}^\rightharpoonup}^2 + g}}}
+ u^r \partial_r \mathop{x}^\rightharpoonup \cr
\dot{\mathop{p}^\rightharpoonup} &
= \partial_r \left( {{gg^{rs}\partial_s
\displaystyle\mathop{x}^\rightharpoonup}%
\over{\sqrt{ \displaystyle{\mathop{p}^\rightharpoonup}^2 + g}}}
+ u^r \mathop{p}^\rightharpoonup \right) \cr
}~~,
\eqno{(16)}
$$
from which $u^r$ can be determined as
$$
u^r = g^{rs}
\dot{\mathop{x}^\rightharpoonup} \partial_s
\mathop{x}^\rightharpoonup~~.
\eqno{(17)}
$$
(16)/(17) are equivalent to the Lagrangian equations of motion,
$$
\partial_t \big( {{\delta {\cal L}}\over{\delta \dot{x}^i}} \big)
+ \partial_r \big( {{\delta {\cal L}}\over{\delta (\partial_r x^i)}}
\big)
= 0~~.
\eqno{(18)}
$$
The effect of choosing the timedependence of
$ \varphi = ( \varphi' \cdots \varphi^M)$
such that
$\displaystyle \dot{\mathop{x}^\rightharpoonup} \partial_r
\displaystyle\mathop{x}^\rightharpoonup \equiv 0 $
is therefore not (6), but putting $ u^r =0$, in (15).

Note that (5) and (6) are therefore valid for arbitrary codimension
(i.e. $M$-\break
dimensional extended objects in $D$-dimensional Minkowski-
space).

In any case, the question is how to proceed ( from (5)/(6) ).
At first, one may hope that the existence of $n$ time-independent,
$\varphi$ dependent functionals ( of the $n$ fields
$\displaystyle\mathop{x}^\rightharpoonup$  and their conjugate momenta,
$\displaystyle\mathop{p}^\rightharpoonup$ )
will be sufficient to have some kind of 'infinite-dimensional Liouville
integrability'.
However, whereas ${\cal H}(\varphi)$ commutes ( weakly ) with itself,
$$
\left\{
\int_\Sigma f(\varphi) {\cal H}(\varphi),
{}~\int_\Sigma h(\tilde\varphi) {\cal H}(\tilde\varphi) \right\}
=
\int_\Sigma (f \partial_r h - h \partial_r f) gg^{rs}~
{{C_s}\over{\cal H}}~~,
\eqno{(19)}
$$
it does not commute with $C_r$:
$$
\left\{
{\cal H}(\varphi),~C_r(\tilde\varphi) \right\}
\approx \partial_r \delta ^{(M)} (\varphi,~\tilde\varphi)~~.
\eqno{(20)}
$$
One may try to subtract from $C_r$  a term $\partial_r Y$, $Y$
conjugate to ${\cal H}$, or
enlarge the phase-space by a pair of conjugate fields, or argue, that
the ( weak ) commutativity of ${\cal H}$ with itself already provides a
separation of variables for the extended object ( 'up to projecting
onto
$C_r = 0'$ ).
However, if one performs the Hamiltonian reduction by choosing, e.g.,
$$
\Pi ^r: = \varphi^r - x^r \equiv 0~, \hbox to 3mm{}r = 1,~\cdots,~M
\eqno{(21)}
$$
( strictly speaking, this way of gauge-fixing is globally possible only
for certain infinitely extended surfaces ) one would have
$\{ C_r,~\Pi ^s \}$ $= \partial_r x^s \approx \delta _r^s$, hence
$$
\eqalign{
\{ F,~G \}^* &
= \{ F,~G\} - \int_\Sigma \{ F,~C_r(\varphi)\}~
\{\Pi ^r(\varphi),~G \} d^M \varphi \cr
 &
+ \int_\Sigma \{ G,~C_r(\varphi) \}~
\{ \Pi ^r\varphi),~F\} d^M \varphi \cr
}
\eqno{(22)}
$$
for the Dirac-bracket on the reduced phase-space, and due to (20),
${\cal H}$ will no longer be conserved, as
$$
\mathop{{\cal H}}^\circ : =
\{ {\cal H}, ~H \}^* \approx \partial_r ( p_a \partial_r x_a )
\not= 0~~, \hbox to 3mm{}a = M+1, ~\cdots,~n~~.
\eqno{(23)}
$$
The left-over fields,however, remain conjugate in the reduced
Hamiltonian,
$$
H = \int \sqrt{p_a p_a + p_a p_b \partial_r x_a \partial_r x_b
+ det ( \delta _{rs} + \partial_r x_a \partial_s x_a)} ~~,
\eqno{(24)}
$$
which for the hypersurface case simplifies $(z=x^n)$ to
$$
H = \int \sqrt{1+p^2}~\sqrt{1+\partial_r z \partial_r z}~~,
\eqno{(25)}
$$
and agrees, in the case of membranes, with the one derived,
(in a rather different way) in [3].
A more interesting way to fix the gauge is to use ${\cal H}$ itself,
by demanding e.g.
$$
\Pi : = {\cal H} - \rho (\varphi) H \equiv 0~~, \hbox to 3mm{}
\int_\Sigma \rho (\varphi) = 1 ~~.
\eqno{(26)}
$$
When following the Dirac-procedure [6], it is probably best to split
(6)
into\break
$C =  \partial_r$  $C_r  = 0$, and the complement
(the elimination of $C$ and $\Pi $ will then involve the Green's
function
for the Laplacian on $\Sigma$ )
However, one may also work with the symplectic form
$\int_\Sigma d \displaystyle\mathop{x}^\rightharpoonup (\varphi)$
$\wedge d \displaystyle\mathop{p}^\rightharpoonup(\varphi)$
in the following way:
Restricting to the hyper-surface case, one first solves (6) by
$$
\eqalign{
& \mathop{p}^\rightharpoonup =
p \cdot \mathop{m}^\rightharpoonup \cr
& M~\cdot~(\mathop{m}^\rightharpoonup)_i =
\epsilon _{ii_1 \cdots i_M} ~\epsilon ^{r_1\cdots r_M} ~\partial_{r_1}
x^{i_1}
\cdots ~\partial_{r_M} ~x^{i_M} \cr
}
\eqno{(27)}
$$
( at this point one should perhaps note the existence of a somewhat
special one-parameter class of canonical transformations,
$$
(\mathop{x}^\rightharpoonup,~\mathop{p}^\rightharpoonup ) \rightarrow
({\mathop{x}^\rightharpoonup}^\lambda  :~
= \mathop{x}^\rightharpoonup,~{\mathop{p}^\rightharpoonup}^\lambda :~
= \mathop{p}^\rightharpoonup + \lambda \mathop{m}^\rightharpoonup)~~,
\eqno{(28)}
$$
generated by
$$
Q_\lambda  = {1 \over {(M+1)}}~\int_\Sigma \lambda
\mathop{x}^\rightharpoonup
\cdot \mathop{m}^\rightharpoonup
 \eqno{(29)}
$$
- which allows e.g. to express $H$ as
$\displaystyle{1 \over \sqrt{2}}\int_\Sigma$
$\sqrt{\displaystyle{\mathop{p}^\rightharpoonup}^2_+
+ \displaystyle{\mathop{p}^\rightharpoonup}^2_-}$ ).
In any case, upon (27) the symplectic form becomes
$$
- \int_\Sigma \epsilon ^{r_1 \cdots r_M} \epsilon _{ii_1 \cdots i_M}
(\partial_{r_1} x^{i_1} \cdots \partial_{r_M} x^{i_M}
{{dp \wedge dx^i}\over{M}}
+ \partial_r p \partial_{r_2} x^{i_2} \cdots \partial_{r_M} x^{i_M}
dx^i \wedge dx^{i_1}~~.
\eqno{(30)}
$$
Changing variables from ($\displaystyle\mathop{p}^\rightharpoonup$,
$\displaystyle\mathop{x}^\rightharpoonup$) to (${\cal
H}=\sqrt{p^2+1}\sqrt{g}$,
$\displaystyle\mathop{x}^\rightharpoonup$), and then
eliminating part of the degeneracy of (30) by using (26),
(30) will be of the form
$$
\int_\Sigma w_i (\varphi) dE\wedge dx^i(\varphi) +
\int_{\Sigma \times \Sigma} w_{ij}(\varphi,~\tilde\varphi)~
dx^i(\varphi)\wedge dx^j(\tilde\varphi)~~.
\eqno{(31)}
$$
After eliminating the remaining $1 + (M-1)\cdot \infty $
degenerate directions in (31),
one must find $Q$ and $P$ ( in terms of $E$ and
$\displaystyle\mathop{x}^\rightharpoonup$  ) such that, at least
locally,
$$
w = d Q \wedge d P ~~.
\eqno{(32)}
$$
The dynamics will then be given by
$$
E = E(Q,~P)~~.
\eqno{(33)}
$$
Even for the string case ( closed strings in 3-dimensional flat
Minkowski
space ) this point of view should be quite interesting ( while it seems
still difficult to find Q and P, one knows that there must exist
infinitely many conserved quantities for (33) ). In this case,
$\{ \Pi (\varphi),~C(\tilde\varphi)\}$
$\approx \delta ' (\varphi - \tilde\varphi)$
$= : \chi (\varphi,~\tilde\varphi)$.
Apart from having to take proper care of the zero eigenvalue of $\chi $
the difference of the Dirac- and the original bracket will be
$$
\int_{\Sigma \times \Sigma}~\{~\cdot,~{\cal H}(\varphi)- H\rho
(\varphi) \}~
\theta (\varphi,~\tilde\varphi)~\{ C(\tilde\varphi),~\cdot~\}
\eqno{(34)}
$$
( antisymmetrized ). However, as both $C$ and {\cal H} commute with
$H$,
the time-evolution of $x_1$  and $x_2$ will be unaltered :
$$
\dot{x}_r = p_r = p\epsilon _{rs}x'_s =
\sqrt{{{1}\over{\displaystyle{\mathop{x}^\rightharpoonup}'^2}}-
{{1}\over{E^2\rho ^2}}}
{}~~\epsilon _{rs} x'_s~\hbox to 5mm{}~'=
{\partial\over{\partial\varphi}}~~.
\eqno{(35)}
$$
In the membrane-case one will get the generalized $su(\infty )$ Nahm
equations [3] this way.
Somewhere in between ( in complexity) are axially symmetric
membranes ( for which a zero-curvature-condition was given in [4], and
derived to be equivalent to strings in a curved 3-dimensional space, in
[7]), with equations of motion
$$
\eqalign{
\dot{r} = & ~\sqrt{ { 1 \over g } - {1 \over {E^2 \rho
^2}}}~~z'~\cdot~r \cr
\dot{z} = & -\sqrt{ { 1 \over g } - {1 \over {E^2 \rho
^2}}}~~r'~\cdot~r \hbox to 7mm{}
(g=r^2 ( r'^2  + z'^2) )~~. \cr
}
\eqno{(36)}
$$

\vskip7mm
\noindent
Acknowledgement:

I benefited from conversations with many members and visitors at
Yukawa Institute, the Research Institute for Mathematical Sciences,
and Kyoto University. In particular, I would like to thank T.Inami
and K.Takasaki, - as well as Yukawa Institute for hospitality. I am
very grateful to K. Hayashi for typing this paper, and to the
Deutsche Forschungsgemeinschaft for financial support.

\endpage
\noindent
References:

\item{[1]} J.Hoppe; 'Quantum Theory of a Massless Relativistic Surface
...'
 MIT Ph.D.-thesis and Elem. Part. Res. J. (Kyoto) {\bf 80} (1989) 145.
\item{[2]} see e.g. 'Supermembranes and Physics in 2+1 Dimensions'
Proceedings of the 1989 Trieste Conference, World Scientific 1990,
Eds. M. Duff, C. Pope, E. Sezgin.
\item{[3]} M.Bordemann, J.Hoppe; Phys. Lett. {\bf B 325} (1994) 359.
\item{[4]} J.Hoppe; 'Surface Motions and Fluid Dynamics'
Phys. Lett. {\bf B} (to appear).
\item{[5]} A.Ashtekar; 'Lectures on Non-Perturbative Canonical Gravity'
World Scientific 1991.
\item{[6]} P.A.M.Dirac; 'Lectures on Quantum Mechanics' Academic Press
1967.
\item{[7]} J.Hoppe; Phys. Lett. {\bf B 329} ( 1994 ) 10.

\bye